\documentclass[english,aps, pra, eqsecnum, preprint]{paper}
\usepackage[T1]{fontenc}
\usepackage[latin9]{inputenc}
\usepackage{bm}
\usepackage{amsmath}
\usepackage{graphicx}
\usepackage[numbers]{natbib}

\makeatletter
\newcommand{\lyxaddress}[1]{
	\par {\raggedright #1
	\vspace{1.4em}
	\noindent\par}
}

\makeatother

\usepackage{babel}
\begin{document}
\title{Initial states and apodisation for quantum field simulations in phase-space}
\author{Peter D. Drummond}
\maketitle

\lyxaddress{Swinburne University of Technology, Melbourne, Victoria 3122, Australia\\
 JILA, University of Colorado and National Institute of Standards
and Technology, Boulder, CO 80309, USA.}
\begin{abstract}
Bosonic quantum fields can be simulated with ``quantum software''
in phase-space. The positive-P, Wigner and Q-function phase-space
methods are reviewed. Initial quantum states and boundaries for infinite
domains are considered in detail. The quantum initial conditions treated
include both multi-mode Gaussian states and number states, which are
sampled using phase-shifted weighting techniques. This is more efficient
than direct probabilistic sampling. Algorithms for treating periodic
boundaries within infinite domains are developed via apodisation,
or absorbers near the boundary. Similar truncation or anti-aliasing
issues arise in the frequency domain with Fourier transform methods.
Complex apodisation, in which there are phase-shifts as well as absorption,
improves accuracy by reducing unwanted errors, and this is illustrated
numerically. The complex power law apodisation techniques described
here can be applied to numerical calculations or even directly in
optics. Phase-space simulations, used for both quantum optical and
Bose-Einstein condensate quantum dynamical calculations, require quantum
noise terms to be included. A method is described to analyze number
conservation in apodised calculations. Some techniques developed here
are potentially applicable to all digital simulations, whether using
classical or quantum logic.
\end{abstract}

\section{Introduction}

Phase-space methods are widely used in quantum simulations. Wigner
\citep{Wigner_1932} developed the first quantum phase-space method.
Husimi \citep{Husimi1940} invented a positive representation, and
later Moyal \citep{Moyal_1949} derived a dynamical theory. Glauber
adapted these approaches to normally-ordered operators in laser physics
\citep{Glauber_1963_P-Rep,Cahill_Galuber_1969_Density_operators}.
These techniques were later extended, shown to be equivalent to stochastic
processes \citep{Drummond_Gardiner_PositivePRep} and used to simulate
quantum fields in super-fluorescence, pulsed quantum squeezing \citep{Carter:1987,DrummondEberly1981PhysRevA.25.3446},
and many other quantum processes \citep{Drummond:2016}. 

This paper describes essential components of space-time quantum field
dynamical simulations. These have been used to treat very large Hilbert
spaces, up to to millions of qubit equivalents \citep{Deuar:2007_BECCollisions}.
In order to utilize such techniques one must both generate initial
quantum states, and implement boundary conditions. Gaussian initial
states are relatively simple to implement. The important case of initial
number states is also treated here. For both initial and boundary
conditions, techniques using complex weights and phase-shifts are
shown to be effective.

Fast Fourier transform (FFT) techniques \citep{cooley1965algorithm}
are frequently used with phase-space representations to numerically
propagate quantum fields. Due to their accuracy and speed, FFTs are
used in classical optics, laser physics, photonics and BEC calculations.
The use of such methods, especially in infinite domains, requires
the treatment of truncation in ordinary and momentum space. Other
methods of integration, with non-uniform grids and finite differences
or related spectral techniques, are also possible \citep{Rooney2014}.
These alternatives do not alleviate the fundamental issue, common
to all field simulations, which is the problem of representing a large
space-time volume with a finite number of lattice points.

FFT techniques usually employ periodic boundary conditions, which
permit an efficient solution in quantum field simulations. Thermal
and quantum noise can be easily included. However, periodic boundary
conditions can cause errors. This is especially an issue with localized
initial fields on infinite backgrounds, where a periodic boundary
condition is not physical. The fields can propagate through the boundaries
and wrap around. There are similar ``wrap-around'' or aliasing errors
in frequency space. This is a generic problem, and holds not just
in the case of transverse diffraction. It also occurs in the case
of short pulse propagation, where dispersion effects play a similar
role to diffraction. Again, it is usually most efficient to use Fourier
transform techniques on a finite window, with periodic boundary conditions.
We regard temporal windowing as a space window, just as in the case
of transverse diffraction effects.

On long time-scales, periodic boundary conditions in space lead to
wrap-around that is not present in the target system. Similar issues
also arise in optical instruments of many kinds, where a hard boundary
causes unwanted diffraction and interference \citep{JACQUINOT196429}.
While such devices do not have periodic boundaries, the use of gradual
absorption or apodisation has long been used to alleviate problems
due to diffraction at hard boundaries. In the software world, boundary
absorption or apodisation is a useful way to prevent boundary artifacts
in either momentum or space \citep{orszag1971,DRUMMOND1983211}. 

For quantum software using phase-space methods, when simulating quantum
fields, one must also consider the effects of apodisation on noise
and conservation laws. Without taking account of this, errors can
occur in the field commutation relations and in calculating the total
numbers of particles. Any proposal must satisfy the requirement that
in a some neighborhood of $\bm{x}=0$, there is a minimal perturbation
of the exact quantum solution. 

This paper treats both quantum initial conditions and apodisation
methods. In both cases, complex weights and phase-shifts are utilized
to improve simulation efficiency. Related techniques are also used
in lithography and lens design \citep{levenson1982improving}. Here
we show how such ideas can be used to simulate quantum field dynamics
with a finite digital machine. This issue is likely to occur in all
types of quantum simulations, including possible future quantum computers. 

\section{Phase-space representations}

We start by summarizing the properties of phase-space representations.
The positive-P and truncated Wigner-Moyal phase-space methods have
been employed to simulate one-dimensional (1D) quantum field dynamics
in photonics \citep{Drummond:1993EPL}, and in Bose-Einstein condensate
(BEC) quantum dynamics in higher dimensions \citep{Steel1998}. Both
methods were also applied to full simulations of optical fibers, including
Raman effects \citep{Drummond:2001a}.

Later applications included large-scale truncated Wigner simulations
of BEC collisions \citep{Norrie:2005}, simulated quantum transport
in a BEC in an optical lattice \citep{Ruostekoski2005}, detailed
comparison of experiment and theory for optical fiber, to below the
quantum noise level \citep{Corney:2006_ManyBodyQD}, comparisons of
3D positive-P and Wigner simulations of BEC collisions with 150,000
atoms from first principles \citep{Deuar:2007_BECCollisions}, proposals
for atomic Hong-Ou-Mandel correlations from BEC collisions \citep{lewis2016proposal},
violations of Bell inequalities \citep{DrummondOPan2014,RosalesZarateBellPhysRevA.90.022109},
and many other applications \citep{Drummond:2016}.

The starting point is an $M-$mode Hilbert space with a number-state
basis of $\left|\bm{n}\right\rangle $, where $\bm{n}=\left(n_{1,}\ldots n_{M}\right)$.
The Bose operators are $\hat{\bm{a}}$, where $\hat{\bm{a}}=\left(\hat{a}_{1,}\ldots\hat{a}_{M}\right)$.
Extended vectors are defined as $\overrightarrow{\bm{a}}=\left[\hat{\bm{a}},\hat{\bm{a}}^{\dagger}\right]$,
so that $\hat{a}_{j}^{\dagger}=\hat{a}_{j+M}$. There is a progression
of phase-space representations from normally-ordered, to symmetrically
ordered, then anti-normally-ordered approaches. The corresponding
phase-space distributions have an increasing variance as one goes
from normal to anti-normal ordering.

\subsection{P-representation}

The Glauber P-representation \citep{Glauber_1963_P-Rep} uses a diagonal
coherent projector, and is singular for non-classical fields. The
generalized P-representation \citep{Drummond_Gardiner_PositivePRep}
extends this by using an off-diagonal coherent projector to treat
all quantum states without singularities. It is defined as a real
or complex phase-space function $P(\vec{\alpha})$, such that:

\begin{equation}
\hat{\rho}=\int P(\vec{\alpha})\hat{\Lambda}\left(\vec{\alpha}\right)\mathrm{d\mu\left(\vec{\alpha}\right)}\,.
\end{equation}
Here $\vec{\alpha}=(\bm{\alpha},\bm{\beta})$ is a complex vector
in a non-classical phase-space of $4M$ real phase-space dimensions,
and $\mathrm{d\mu\left(\vec{\alpha}\right)}$ is an integration measure.
This can either be a volume or surface measure in the enlarged phase-space
$\vec{\alpha}$. The result is an exact expansion of the quantum density
matrix in terms of the coherent state projection operator 
\begin{equation}
\hat{\Lambda}\left(\vec{\alpha}\right)=\frac{\left|\bm{\alpha}\right\rangle \left\langle \bm{\beta}^{*}\right|}{\left\langle \bm{\beta}^{*}\right|\left.\bm{\alpha}\right\rangle }=\prod_{k}\hat{\Lambda}_{k}\left(\alpha_{k},\beta_{k}\right)\,.
\end{equation}
This projector is defined in terms of $M$-mode Bargmann-Glauber coherent
states $\left\Vert \bm{\alpha}\right\rangle $, where
\begin{equation}
\left|\bm{\alpha}\right\rangle \equiv\prod_{k=1}^{M}\left[e^{-\left|\alpha_{k}\right|^{2}/2}\sum_{n_{k}}\frac{\alpha_{k}^{n_{k}}}{\sqrt{n_{k}!}}\left|n_{k}\right\rangle \right]\,.
\end{equation}

A consequence of the definition is that the P-function is normalized,
which follows since $Tr\left(\hat{\rho}\right)=Tr\left(\hat{\Lambda}\right)=1$,
so that $\int P(\vec{\alpha})\mathrm{d\mu\left(\vec{\alpha}\right)=1}$.
There are existence theorems and explicit constructions for all quantum
states, known for both complex and positive forms. The most general
case is a complex $P(\vec{\alpha})$, written as 
\begin{equation}
P(\vec{\alpha})=\Omega(\vec{\alpha})P_{+}(\vec{\alpha})\,,
\end{equation}
where $P_{+}(\vec{\alpha})$ is real, positive and normalized, so
it can be sampled with conventional probability techniques:
\begin{equation}
\int P_{+}\left(\vec{\alpha}\right)\mathrm{d\mu\left(\vec{\alpha}\right)=1}.
\end{equation}

Here $\Omega(\vec{\alpha})$ is a complex weight, which is ideally
chosen so that has a nearly uniform modulus, $\left|\Omega(\vec{\alpha})\right|\sim1$,
to give efficient sampling. From these definitions, it has a mean
value of unity:
\begin{equation}
\left\langle \Omega\left(\vec{\alpha}\right)\right\rangle _{+}\equiv\int\Omega\left(\vec{\alpha}\right)P_{+}\left(\vec{\alpha}\right)\mathrm{d\mu\left(\vec{\alpha}\right)=1}\,.
\end{equation}

It is also convenient to define a weighted mean, which includes the
complex weights, so that:
\begin{equation}
\left\langle O(\vec{\alpha})\right\rangle _{P}\equiv\int\int O(\vec{\alpha})P(\vec{\alpha})\mathrm{d\mu\left(\vec{\alpha}\right)}.
\end{equation}

Averages in P-distributions correspond quantum-mechanically to normal-ordered
quantum averages, denoted $:\ldots:$. These have creation operators
like $\hat{\bm{a}}^{\dagger}$ ordered to the left, and if $O(\vec{\alpha})$
is a power series in $\vec{\alpha}$, then the following equivalence
holds:
\begin{equation}
\left\langle O(\vec{\alpha})\right\rangle _{P}=\left\langle :O(\hat{\bm{a}},\hat{\bm{a}}^{\dagger}):\right\rangle \,.
\end{equation}

The average $\left\langle \hat{O}\right\rangle \equiv Tr\left(\hat{O}\hat{\rho}\right)$
is a standard quantum ensemble average with density matrix $\hat{\rho}$,
and $\left\langle O\right\rangle _{P}$ indicates the P-function average.
When sampling the distribution with $S$ samples, averages are obtained
from samples of $\vec{\alpha}^{(n)}$ distributed with probability
$P_{+}(\vec{\alpha}^{(n)}),$so that:

\begin{equation}
\left\langle O(\vec{\alpha})\right\rangle _{P}=\lim_{S\rightarrow\infty}\frac{1}{S}\sum_{n=1}^{S}\Omega(\vec{\alpha}^{(n)})O(\vec{\alpha}^{(n)})\,.
\end{equation}

The main drawback of this approach is that the sampling error grows
in time due to the enlarged phase-space, and in some cases there are
boundary term errors in phase-space. This requires the use of additional
stabilizing gauges, and limits the time-duration that can be usefully
simulated \citep{Gilchrist_Gardiner_PD_PPR_Application_Validity,Deuar:2001,Deuar2006a,Deuar2006b}.

\subsection{Wigner representation}

Other representations are known, including the Wigner and Q-representations.
These correspond to symmetrically ordered and anti-normally ordered
operators respectively. The phase-space is classical, with $\vec{\alpha}\equiv\left[\bm{\alpha},\bm{\alpha}^{*}\right]$.
They can be written formally as convolutions of the generalized P-function,
extending a result of Cahill and Glauber \citep{Cahill_Galuber_1969_Density_operators}:

\begin{align}
P_{s}\left(\vec{\alpha}\right) & =\left(\frac{1}{\pi s}\right)^{M}\int P(\vec{\alpha}_{0})e^{-\left(\bm{\alpha}-\bm{\alpha}_{0}\right)\cdot\left(\bm{\alpha}^{*}-\bm{\beta}_{0}\right)/s}\mathrm{d\mu\left(\vec{\alpha}_{0}\right)}\,\nonumber \\
 & =\left(\frac{1}{\pi s}\right)^{M}\left\langle e^{-\left(\bm{\alpha}-\bm{\alpha}_{0}\right)\cdot\left(\bm{\alpha}^{*}-\bm{\beta}_{0}\right)/s}\right\rangle _{P}.\label{eq:convolution-1}
\end{align}

One can define the original Glauber-Sudarshan P-representation as
the singular limit of the convolution in the limit of $s\rightarrow0$.
However, this often may exist only as a generalized function, depending
on the quantum state involved. As a result, we will use the positive-P
distribution $P(\vec{\alpha})$ instead, so that all relevant nonsingular
cases can be treated.

There are multiple ways to define the Wigner representation. One way
to define this is as a convolution of the generalized P-function as
given above. Thus, a delta-function P-representation representing
a coherent state corresponds to a Gaussian Wigner function.

For the case of the Wigner function one has $s=1/2$, so that $P_{1/2}\left(\vec{\alpha}\right)\equiv W\left(\vec{\alpha}\right)$.
The Wigner function is always nonsingular and real. This representation
is also generally non-positive, making direct probabilistic sampling
impossible except for the special cases described below, or by using
weighted sampling methods with non-positive weights.

Averages are defined as integrals over the Wigner function, so if
$d\vec{\alpha}$ is now the complex phase-space volume integral, corresponding
to $2M$ real dimensions, then:
\begin{equation}
\left\langle O(\vec{\alpha})\right\rangle _{W}\equiv\int\int O(\vec{\alpha})W(\vec{\alpha})d\vec{\alpha}.
\end{equation}
Wigner phase-space averages like this correspond quantum-mechanically
to symmetrically-ordered quantum averages. If we define $\left\{ O(\hat{\bm{a}},\hat{\bm{a}}^{\dagger})\right\} _{s}$
as an s-ordered average, which could be normally ordered, symmetrically
ordered or anti-normally ordered, then:
\begin{equation}
\left\langle O(\vec{\alpha})\right\rangle _{W}=\left\langle \left\{ O(\hat{\bm{a}},\hat{\bm{a}}^{\dagger})\right\} _{1/2}\right\rangle _{Q}.
\end{equation}
Just as with the P-function, one can also define a weighted version,
so that, if $W_{+}(\vec{\alpha})$ is defined as a positive probability,
then:
\begin{equation}
W(\vec{\alpha})=\Omega(\vec{\alpha})W_{+}(\vec{\alpha}).
\end{equation}

Other methods for direct sampling are available in the cases where
$W\left(\bm{\alpha}\right)$ is positive, but for pure states these
are always Gaussian cases, unless approximations are used. The double-dimension
formalism of the positive-P method is also applicable here, if we
define a positive Wigner function on an extended phase-space $\vec{\alpha}=\left[\bm{\alpha},\bm{\beta}\right]$
\citep{hoffmann2008hybrid}. 

Typically, the Wigner dynamical equations are truncated using an $1/N$
expansion for $N$ bosons, in order to obtain tractable stochastic
equations.

\subsection{Q-function}

The Q-function, originally proposed by Husimi \citep{Husimi1940},
is defined either using the convolution method, or directly using
the fundamental definition that:
\begin{equation}
Q(\vec{\alpha})=\frac{1}{\pi^{M}}\left\langle \bm{\alpha}\right|\hat{\rho}\left|\bm{\alpha}\right\rangle \,.
\end{equation}

As with the Wigner case, the extended vector $\vec{\alpha}$ is also
applicable here if one defines $\vec{\alpha}=\left[\bm{\alpha},\bm{\alpha}^{*}\right]$.
For the case of the Q-function one has $s=1$, so that $P_{1}\left(\vec{\alpha}\right)\equiv Q\left(\vec{\alpha}\right)$.
The Q-function is always nonsingular, real and positive. Direct probabilistic
sampling is therefore possible. This method is also applicable to
qubits and fermions \citep{Arecchi_SUN,FermiQ}, and has been used
for simulating large-order correlations and Bell violations \citep{ReidqubitPhysRevA.90.012111,Opanchuk:2014}.

Averages are defined as integrals over the Q-function, so if $d\vec{\alpha}$
is the complex phase-space volume integral, corresponding to $2M$
real dimensions, then:
\begin{equation}
\left\langle O(\vec{\alpha})\right\rangle _{Q}\equiv\int\int O(\vec{\alpha})Q(\vec{\alpha})\mathrm{d\vec{\alpha}}.
\end{equation}
Averages like this correspond quantum-mechanically to anti-normally-ordered
quantum averages, ie,
\begin{equation}
\left\langle O(\vec{\alpha})\right\rangle _{Q}=\left\langle \left\{ O(\hat{\bm{a}},\hat{\bm{a}}^{\dagger})\right\} _{1}\right\rangle _{Q}\,.
\end{equation}

While the Q function is always positive, it does not generally have
a normal stochastic process. Instead, it corresponds to a stochastic
process that propagates simultaneously forward and backwards in time
\citep{drummond2019Q,drummond2019time}, requiring more specialized
techniques to solve it. 

\section{Sampling initial quantum states}

To treat large quantum systems, there is a need for computational
scalability as the number of modes increases. This means that probabilistic
sampling is necessary, as direct high-dimensional integration over
a multimode distribution is exponentially complex in the number of
modes. Sampling is essential for computability purposes. 

Scalability is even an issue when there are exact solutions for the
eigenstates, because representing an arbitrary initial quantum state
requires a superposition of exponentially many eigenstates, if one
has a large number of quantum particles or modes. In practice, this
is a severe limitation on dynamical calculations even in exactly soluble
models \citep{yurovsky2017dissociation}.

As a result, computability requires an efficient strategy for sampling
a range of suitable initial conditions.

\subsection{Gaussian states}

It is straightforward to sample the phase-space distributions described
above if the initial conditions are Gaussian states. Examples of these
include coherent states \citep{Glauber_1963_P-Rep}, squeezed states
\citep{Drummond2004_book}, Bogoliubov states \citep{Ruostekoski2005,Ruostekoski2010,lun2019phase},
and thermal states. In these cases, sampling can be obtained by generating
$2M$ real Gaussian noises defined so that 
\begin{equation}
\left\langle w_{i}w_{j}\right\rangle =\delta_{ij},
\end{equation}
together with corresponding random phase-space samples where $\vec{\alpha}=\left[\alpha_{1},\ldots\alpha_{2M}\right]$
, such that: 
\begin{equation}
\alpha_{i}=\alpha_{i}^{0}+B_{ij}w_{j}.
\end{equation}
Here the quantum mean value is given by:

\begin{equation}
\left\langle \hat{a}_{i}\right\rangle _{Q}=\alpha_{i}^{0}.
\end{equation}

In order to achieve an s-ordered cross-correlation - including the
entire extended vector of $2M$ annihilation and creation operators
- of
\begin{equation}
\left\langle \left\{ \hat{a}_{i}\hat{a}_{j}\right\} _{s}\right\rangle _{Q}=\alpha_{i}^{0}\alpha_{j}^{0}+\sigma_{ij},
\end{equation}
 one simply chooses the $2M\times2M$ complex square root $\mathbf{B}$,
such that:
\begin{equation}
\sigma_{ij}=B_{ik}B_{jk}.
\end{equation}

There are restrictions on $\bm{B}$ for the Wigner and Q-functions.
The condition that $\alpha_{j}^{*}=\alpha_{M+j}$ implies that: 
\begin{equation}
B_{ik}^{*}=B_{ik+M}.
\end{equation}
In some ``quasi-Gaussian'' states, like a laser or a BEC, the true
quantum states have no well-defined phase \citep{Louisell}. In such
cases, one can generate a Gaussian state relative to some phase-reference
amplitude, then randomize the phase. If the dynamics and the final
measurements are phase-independent, as is often the case, the last
step of phase-averaging is not necessary, as it doesn't change the
results. In these cases a coherent state is equivalent to a Poissonian
average over number states, and therefore either type of input can
be used \citep{lun2019phase}. 

\section{Number state sampling}

In some cases, one wishes to sample the phase-space for a number state,
or a superposition of number states. These are more difficult to sample
than Gaussian states, but certainly not impossible. The most general
possible quantum states can be characterized by their matrix elements
in a number state basis such that:
\begin{equation}
C_{\bm{nm}}=\left\langle \bm{n}\right|\hat{\rho}\left|\bm{m}\right\rangle ,
\end{equation}
or alternatively, that:
\begin{equation}
\hat{\rho}=\sum_{\bm{nm}}C_{\bm{nm}}\left|\bm{n}\right\rangle \left\langle \bm{m}\right|.
\end{equation}
For a number state mixture, one would have that all the off-diagonal
terms vanish, and:
\begin{equation}
\hat{\rho}=\sum_{\bm{n}}C_{\bm{n}}\left|\bm{n}\right\rangle \left\langle \bm{n}\right|.
\end{equation}
We treat examples of these in this section.

\subsection{P-function for number state expansions}

While the positive P-distribution $P(\vec{\alpha})$ always exists,
this is not always the most numerically efficient choice for sampling.
While one can always construct a number state using the positive P-representation
\citep{Olsen:2009}, there are other techniques that are more effective
in terms of sampling. Choosing the complex P-representation adds a
complex weight to the initial condition, while still allowing the
usual propagation equations for subsequent time evolution. 

This method has been used in simulations of boson sampling equivalent
to permanents as large as $100\times100$ \citep{drummond2016scaling,opanchuk2018simulating,opanchuk2019robustness},
with sampling errors that scale better than experiments. It has also
been applied to simulations of BEC quantum dynamics in one dimension,
with initial quantum number states \citep{ng2019nonlocal}. This technique
was described previously, and we review it in greater detail here.

We focus on the complex P-distribution, which is more compact than
the positive P-distribution \citep{Drummond_Gardiner_PositivePRep}.
The existence theorem states that, for an expansion of an arbitrary
density matrix in number states such that $C_{\bm{nm}}=\left\langle \bm{n}\right|\hat{\rho}\left|\bm{m}\right\rangle $,
and for circular contours enclosing the origin in each amplitude,
then by Cauchy's theorem,
\begin{equation}
P(\vec{\alpha})=\left(\frac{-1}{4\pi^{2}}\right)^{M}e^{\bm{\alpha}\cdot\bm{\beta}}\sum_{\bm{n},\bm{m}}C_{\bm{nm}}\prod_{kl}\frac{\sqrt{n_{k}!m_{l}!}}{\alpha_{k}^{n_{k}+1}\beta_{k}^{m_{k}+1}}\,.
\end{equation}

For vacuum states, the P-function sample is simply one of $\alpha_{k}=\beta_{k}=0$.
For non-vacuum inputs we use a circular contour of radius $r_{k}$
and make a transition to polar coordinates, so that $\alpha_{k}=r_{k}\exp\left(\mathrm{i}\phi_{k}^{(\alpha)}\right)$
, $\beta_{k}=r_{k}\exp\left(-\mathrm{i}\phi_{k}^{(\beta)}\right)$,
and similarly $d\alpha_{k}=\mathrm{i}\alpha_{k}d\phi_{k}^{(\alpha)}$
and $d\beta_{k}=\mathrm{i}\beta_{k}d\phi_{k}^{(\beta)}$. The radius
$r_{k}$ is chosen to minimize the sampling error, and samples are
generated on the contour, to give a Monte-Carlo sampled contour integral.

While the expansion is always applicable, we now assume for simplicity
that we can choose the mode basis that defines the mode operators
such that the density matrix factorizes with $\hat{\rho}=\prod\hat{\rho}_{k}$.
For multiple independent modes, a generalized P-representation is
defined as:

\begin{equation}
P\left(\bm{\alpha},\bm{\beta}\right)=\prod_{k=1}^{M}P_{k}\left(\alpha_{k},\beta_{k}\right)\,.\label{eq:Complex-P-Dist}
\end{equation}
The phase variables can be understood intuitively on defining: 
\begin{align}
\phi_{k} & =\left(\phi_{k}^{(\alpha)}+\phi_{k}^{(\beta)}\right)/2\nonumber \\
\theta_{k} & =\phi_{k}^{(\alpha)}-\phi_{k}^{(\beta)},
\end{align}
where $\phi_{k}$ is the classical phase, and $\theta_{k}$ is a nonclassical
phase which only exists when the quantum state has nonclassical features.

Hence, one has the result that:
\begin{align}
\alpha_{k}^{-n_{k}} & =e^{-in_{k}\left(\theta_{k}/2+\phi_{k}\right)}r_{k}^{-n_{k}}\nonumber \\
\beta_{k}^{-m_{k}} & =e^{-im_{k}\left(\theta_{k}/2-\phi_{k}\right)}r_{k}^{-n_{k}}.
\end{align}
In this case we get the overall density matrix as:
\begin{equation}
\hat{\rho}_{k}=\sum_{nm}\int_{-\pi}^{\pi}\frac{\mathrm{d}\phi}{2\pi}\int_{-\pi}^{\pi}\mathrm{d}\theta C_{nm}\left(\phi\right)I_{k}^{nm}\left(\theta\right)\hat{\Lambda}_{k}\left(\alpha,\beta\right),
\end{equation}
where we define $C_{nm}\left(\phi_{k}\right)=C_{nm}e^{-\mathrm{i}\left(n_{k}-m_{k}\right)\phi_{k}}$,
and $\bar{n}_{k}=\left(n_{k}+m_{k}\right)/2$, so that $I_{k}^{nm}=0$
for $n_{k}=m_{k}=0$, and otherwise the expansion coefficient $I_{k}^{nm}$
is given by :
\begin{align}
I_{k}^{nm}\left(\theta\right)= & \frac{\sqrt{n!m!}}{2\pi r_{k}^{n+m}}e^{\left(r_{k}^{2}\cos\theta+\mathrm{i}\left(r_{k}^{2}\sin\theta-\bar{n}_{k}\theta\right)\right)}\nonumber \\
 & \,
\end{align}

More general cases with coherences between the modes can be treated,
and lead to additional phase-coherence terms in the contour integral
sampling.

\subsection{Contour sampling techniques}

To illustrate the general sampling procedure, we suppose that each
mode has an input number state, so that if $P_{k}$ is the single-mode
distribution for mode $k$ with $n_{k}$ bosons:

\begin{align}
P_{k}\left(\alpha,\beta\right)=\begin{cases}
\left(\frac{1}{2\pi\mathrm{i}}\right)^{2}\frac{n_{k}!e^{\alpha\beta}}{(\alpha\beta)^{n_{k}+1}} & \mathrm{for}\ n_{k}>0,\\
\delta\left(\alpha\right)\delta\left(\beta\right) & \mathrm{for}\ n_{k}=0.
\end{cases}
\end{align}
After separating the real and imaginary parts of the exponential,
and transforming to phase variables, one can use random probabilistic
sampling for the real part, so that:
\begin{equation}
P_{k}\left(\alpha,\beta\right)=\mathcal{P}_{k}\left(\phi_{k},\theta_{k}\right)\Omega_{k}\left(\phi_{k},\theta_{k}\right).
\end{equation}
We call the distribution $\mathcal{P}_{k}$ a circular von Mises complex-P
distribution (VCP), since the weight on the contour is a von Mises
probability distribution~\citep{BookvonMisesCh3}. Once the phase
angle is randomly chosen,  the imaginary part becomes an additional
complex weight, $\Omega_{k}$. The sampled probability distribution
for the subset of modes with nonzero inputs is then
\begin{align}
\mathcal{P}_{k}\left(\phi_{k},\theta_{k}\right) & =\frac{1}{4\pi^{2}I_{0}\left(r^{2}\right)}\exp\left(r^{2}\cos\theta_{k}\right),\quad n_{k}>0,
\end{align}
where $I_{0}\left(x\right)$ is the modified Bessel function of the
first kind of order $0$, $\theta_{k}\in[-\pi,\pi)$, and $\phi_{k}\in[-\pi,\pi)$.
This corresponds to sampling the variables separately as $\phi_{k}=\mathcal{U}\left(-\pi,\pi\right)$,
and $\theta_{k}=\mathcal{VM}\left(0,r^{2}\right)$, where $\mathcal{U}$
is the uniform distribution, and $\mathcal{VM}$ is the circular von
Mises distribution~\citep{BookvonMisesCh3}. For each sample we calculate
the phase-space variables as $\alpha_{k}=r\exp\left(\mathrm{i}\left(\phi_{k}+\theta_{k}/2\right)\right)$,
$\beta_{k}=r\exp\left(\mathrm{i}\left(\phi_{k}-\theta_{k}/2\right)\right)$.
For vacuum modes where $n_{k}=0$, we use the delta-function distribution
$P_{k}\left(\alpha_{k},\beta_{k}\right)=\delta\left(\alpha_{k}\right)\delta\left(\beta_{k}\right)$;
that is, one takes $\alpha_{k}=\beta_{k}=0$.

The complex weights associated with each sample are

\begin{equation}
\Omega_{k}=\prod_{k=1}^{M}\left\{ \begin{array}{ll}
\frac{\left(n_{k}!\right)I_{0}\left(r^{2}\right)}{r^{2n_{k}}}\exp\left(\mathrm{i}\left(r^{2}\sin\theta_{k}-n_{k}\theta_{k}\right)\right), & n_{k}>0;\\
1, & n_{k}=0.
\end{array}\right.
\end{equation}
These are used to calculate any moment $f(\bm{\alpha},\bm{\beta})$
in conjunction with the $S$ drawn samples of $\bm{\alpha}$ and $\bm{\beta}$
as
\begin{equation}
\langle f(\bm{\alpha},\bm{\beta})\rangle=\frac{1}{S}\sum_{j=1}^{S}\Omega\left(\bm{\alpha}^{(j)},\bm{\beta}^{(j)}\right)f\left(\bm{\alpha}^{(j)},\bm{\beta}^{(j)}\right).
\end{equation}
where $\Omega=\prod\Omega_{k}$ is the total weight associated with
the trajectory.

\subsection{Large radius limit}

It is generally optimal in terms of sampling efficiency to choose
$r_{k}^{2}\approx n_{k}$. For larger particle numbers, these expressions
then simplify, because one can use a large radius for the integration
contour. The phase distribution becomes asymptotically a standard
Gaussian of width $1/r$ in the limit of large $r$, and we use the
asymptotic Bessel function result that:
\begin{equation}
I_{0}\left(z\right)=\frac{e^{z}}{\sqrt{2\pi z}}+O(\frac{1}{z}).
\end{equation}
 The sampled probability distribution for the subset of modes with
nonzero inputs is then a uniform circular distribution of $1/\left(2\pi\right)$,
multiplied by a normalized Gaussian:
\begin{align}
\mathcal{P}_{k}\left(\phi_{k},\theta_{k}\right) & \approx\frac{r}{\left(2\pi\right)^{3/2}}\exp\left(-r^{2}\theta_{k}^{2}/2\right).
\end{align}

In the same large r limit, the weight can be calculated using Stirling's
approximation. Introducing $n!=n^{n}\sqrt{2\pi n}e^{-n}(1+O(1/n))$
and taking $r^{2}=n$, one then obtains for the $k-th$ complex weighting
term:
\begin{align}
\Omega & \approx\frac{n!I_{0}\left(r^{2}\right)}{r^{2n}}\exp\left(\mathrm{i}\left(r^{2}\sin\theta-n\theta\right)\right),\nonumber \\
 & \approx\exp\left(\mathrm{i}\left(r^{2}\sin\theta-n\theta\right)\right).
\end{align}

Therefore it follows that $\Omega=\exp\left(-\mathrm{i}n\theta^{3}/6\right)$
to leading order. Noting that $\left\langle \theta^{2}\right\rangle =1/n$
and hence: $n\theta^{3}\sim1/\sqrt{n}$, these are small phase factors,
and the normalization involved is of order:
\begin{equation}
\left\langle \Omega\right\rangle \sim\left\langle 1-\mathrm{i}n\theta^{3}/6-n^{2}\theta^{6}/72\right\rangle ,
\end{equation}
where one can show that
\begin{equation}
\left\langle \theta^{6}\right\rangle =\frac{15}{n^{3}}.
\end{equation}

As a result, in this approximation the overall weight factor is actually
$\left\langle \Omega\right\rangle \sim1-15/\left(72n\right)$, while
it should be $\left\langle \Omega\right\rangle =1.$ This approximation
error of order $1/n$ can be removed by including higher order terms
of $O(1/n)$, if the next order of accuracy is required.

\subsection{Wigner and Q distribution sampling}

There are several ways to sample the Wigner and Q functions, and we
will describe two possibilities here. The first method is a direct
sampling based on known analytic forms of the function, while the
second method is convolution sampling based on the generalized P-distribution
as a starting point.

One can obtain an analytic form of the Wigner function in terms of
associated Laguerre functions, in the case of a factorized $\hat{\rho}$,
which is diagonal in the number basis, as:
\begin{align}
W\left(\vec{\alpha}\right) & =\left(\frac{2}{\pi}\right)^{M}e^{-2\left|\bm{\alpha}\right|^{2}}\sum_{\bm{n}}C_{\bm{n}}\prod_{k}\left(-1\right)^{n_{k}}\nonumber \\
 & \times\sqrt{\frac{n_{k}!}{m_{k}!}}\left(2\alpha_{k}\right)^{m_{k}-n_{k}}\mathcal{L}_{n_{k}}^{m_{k}-n_{k}}\left(4\left|\alpha_{k}\right|^{2}\right)\,.
\end{align}

From this result, we can see that while the Wigner function always
exists for any quantum state, but it has no generally positive form,
even when obtained from a positive P-function. High-order Laguerre
functions are difficult to calculate numerically, owing to underflow,
overflow and round-off problems. Sampling them is also nontrivial,
as they are extremely oscillatory. As a result, this approach is not
usually employed for carrying out stochastic integration. 

Q-function sampling is more straightforward, as this is always positive.
For a number state $\bm{n}$, one obtains that:
\begin{equation}
Q\left(\vec{\alpha}\right)=\prod_{m}\frac{\left|\alpha_{m}\right|^{2n_{m}}}{\pi n_{m}!}e^{-\left|\alpha_{m}\right|^{2}}.
\end{equation}
This is essentially a product of gamma distributions, for which efficient
sampling methods are known.

\subsection{Convolution sampling}

We will now show that by using a different technique, not based on
Laguerre polynomials, that sampling the full Wigner distribution is
possible, even for relatively complex quantum states. These techniques
have proved useful in establishing the properties of the Wigner function
when a full-scale exact simulation of a quantum Schrodinger Cat transfer
in optomechanics \citep{teh2018creation} was carried out using a
complex P-distribution, which is an exact method.

The results given above for convolutions imply that there is always
a general sampling theorem. Both $Q(\vec{\alpha})$ and $W(\vec{\alpha})$
can always be obtained using sampling methods provided that an algorithm
for sampling the generalized P-function is known, since a convolution
is equivalent to adding a random gaussian variable. In the present
case one must also take into account that the convolution includes
a dimensional reduction from a non-classical phase-space. 

For s-ordered sampling, the fundamental convolution involves a complex
weight of form:
\begin{equation}
e^{-\left(\bm{\alpha}-\bm{\alpha}_{0}\right)\cdot\left(\bm{\alpha}^{*}-\bm{\beta}_{0}\right)/s}.
\end{equation}
To sample either the Wigner or Q-function, we therefore first define
the following linear combinations: 
\begin{align}
\bm{\alpha}_{\pm} & =\frac{1}{2}\left(\bm{\alpha}_{0}\pm\bm{\beta}_{0}^{*}\right).
\end{align}
The variable $\bm{\alpha}_{+}$ is equivalent to a `classical' coordinate,
while $\bm{\alpha}_{-}$ indicates the degree of non-classicality.
To construct a real Gaussian convolution, if we define $\Delta=\bm{\alpha}-\bm{\alpha}_{+}$
and $i\delta=\Delta\cdot\bm{\alpha}_{-}^{*}-\Delta^{*}\cdot\bm{\alpha}_{-}$,
then
\begin{equation}
\left(\bm{\alpha}-\bm{\alpha}_{0}\right)\cdot\left(\bm{\alpha}^{*}-\bm{\beta}_{0}\right)=\left|\Delta\right|^{2}-\left|\bm{\alpha}_{-}\right|^{2}+i\delta.
\end{equation}

The overall distribution is therefore a weighted convolution:

\begin{equation}
W\left(\vec{\alpha}\right)=\left(\frac{1}{\pi s}\right)^{M}\int\int P(\vec{\alpha}_{0})w\left(\vec{\alpha},\vec{\alpha}_{0}\right)e^{-\left|\Delta\right|^{2}/s}\mathrm{d\mu\left(\vec{\alpha}_{0}\right)}
\end{equation}
where the complex weight function is:
\begin{equation}
w\left(\vec{\alpha},\vec{\alpha}_{0}\right)=\mathrm{e^{\left(\left|\bm{\alpha}_{-}\right|^{2}-i\delta\right)/s}.}
\end{equation}
Thus, if one samples the P-function to obtain $\vec{\alpha},$ then
samples the convolution variable, $\Delta_{k}=\Delta_{k}^{x}+i\Delta_{k}^{y}$
as real Gaussian variables with variance $s/2$, such that:
\begin{equation}
\left\langle \left(\Delta_{k}^{x}\right)^{2}\right\rangle =\left\langle \left(\Delta_{k}^{y}\right)^{2}\right\rangle =\frac{s}{2},
\end{equation}
the resulting phase-space coordinate of $\bm{\alpha}=\bm{\alpha}_{+}+\Delta$
is a Wigner or Q-function sample. However, when calculating ensemble
averages with a generalized P-distribution as the starting point,
an additional complex weight must be included with:
\begin{equation}
\Omega_{W}=\Omega_{P}e^{\left(\left|\bm{\alpha}_{-}\right|^{2}-i\delta\right)/s}.
\end{equation}
While this is a complex weight, its average is real. 

\section{Space and momentum boundaries}

After initializing the quantum state, differential identities are
used to obtain dynamical evolution equations in phase-space, which
are described in detail elsewhere \citep{drummond2014quantum}. These
are Fokker-Planck equations in structure, and can either be written
in terms of the mode amplitudes $\vec{\alpha}$, or the corresponding
local fields $\mathbf{\overrightarrow{\psi}}\left(\bm{x}\right)$.
On sampling, the FPEs can be transformed into stochastic partial differential
equations (SPDE). 

When written in Stratonovich \citep{Gardiner_Book_SDE} form, the
field derivative $\mathcal{D}\left[\mathbf{\overrightarrow{\psi}}\right]$
can be expanded as:
\begin{align}
\Delta\mathbf{\overrightarrow{\psi}} & =\mathcal{D}\left[\mathbf{\overrightarrow{\psi}}\right]\Delta t\nonumber \\
 & =\left(\mathbf{A}\left[\mathbf{\overrightarrow{\psi}}\right]+\underline{\mathbf{L}}\left[\bm{\nabla}\right]\cdot\mathbf{\overrightarrow{\psi}}\right)\Delta t+\underline{\mathbf{B}}\left[\mathbf{\overrightarrow{\psi}}\right]\cdot\bm{\Delta w}.
\end{align}

Here, $\mathbf{A}$ is a vector, $\underline{\mathbf{B}}$ a matrix,
$\underline{\mathbf{L}}$ is a linear differential operator in a real
space $\bm{x}$, and $\bm{\Delta w}$ is a Gaussian noise vector,
defined such that:
\begin{eqnarray}
\left\langle \Delta w_{i}\left(\mathbf{x}\right)\Delta w_{j}\left(\mathbf{x}'\right)\right\rangle  & = & \Delta t\delta\left(\bm{x}-\bm{x}'\right)\delta_{ij}\,.
\end{eqnarray}

These equations usually involve truncation or other approximations
in the Wigner case. The situation is different with the Q-function,
which is dynamically equivalent to a forward-backward stochastic equation,
requiring more advanced techniques \citep{drummond2019Q,drummond2019time}. 

SPDEs are often solved using spectral methods, in which the linear
term is converted to an interaction picture \citep{Werner:1997} using
Fourier transforms. Other methods are also possible and are most useful
with trapped BECs \citep{Rooney2014}. Given an initial stochastic
field $\mathbf{\overrightarrow{\psi}}^{(0)}$, a typical midpoint
interaction picture approach using Fourier transforms is to define
an operator $\mathcal{T}=\mathcal{F}^{-1}e^{\underline{\mathbf{L}}\left[i\bm{k}\right]\Delta t/2}\mathcal{F},$which
propagates the amplitude field with the linear operator by $\Delta t/2$,
where $\mathcal{F}$ indicates a forward Fourier transform.

The overall algorithm for a step in time starting at $\mathbf{\overrightarrow{\psi}}^{(0)}$
is therefore a three step process, with a midpoint defined implicitly
as $\mathbf{\overrightarrow{\psi}}^{(m)}=\left(\mathbf{\overrightarrow{\psi}}^{(1)}+\mathbf{\overrightarrow{\psi}}^{(2)}\right)/2$
, and a final result of $\mathbf{\overrightarrow{\psi}}^{(3)}:$
\begin{align}
\mathbf{\overrightarrow{\psi}}^{(1)} & =\mathcal{T}\mathbf{\overrightarrow{\psi}}^{(0)}\nonumber \\
\mathbf{\overrightarrow{\psi}}^{(2)} & =\mathbf{\overrightarrow{\psi}}^{(1)}+\mathbf{A}\left[\mathbf{\overrightarrow{\psi}}^{(m)}\right]\Delta t+\underline{\mathbf{B}}\left[\mathbf{\overrightarrow{\psi}}^{(m)}\right]\cdot\bm{\Delta w}\nonumber \\
\mathbf{\overrightarrow{\psi}}^{(3)} & =\mathcal{T}\mathbf{\overrightarrow{\psi}}^{(2)}.
\end{align}

In calculations of field propagation, apodisation is useful in combination
with fast Fourier transform computational methods. It prevents leakage
through the spatial boundaries from dispersion or diffraction and
similarly prevents leakage in momentum from nonlinear interactions.
This would result in unphysical wrap-around of the fields from the
positive to the negative coordinate axes, in either space or momentum,
due to the use of periodic boundary conditions in space or momentum. 

Both in physical optics and in classical field simulations of such
equations, apodisation means the addition of an artificial absorber
near the boundary. Apodisation in space should be continuous and smoothly
varying in order to reduce diffraction artifacts. This is used to
smooth or filter the field so that it goes to zero smoothly at the
boundary, without additional structure. Because Fourier transforms
are symmetrical between position and momentum, apodisation is required
both in space and momentum \citep{DRUMMOND1983211}. 

Just as dispersion leads to a spreading in position space, nonlinearity
leads to a spreading in momentum space. There are momentum boundaries
at the inverse of the transverse spatial step-size. Such momentum
limits cause aliasing errors, where high-momentum components are wrapped
around to negative or even low momentum. These delocalized jumps caused
by nonlinearities are most effectively de-aliased with projectors
that removes high-momentum terms, to give a final linear translation
$\mathcal{T}_{a}=\mathcal{F}^{-1}e^{\underline{\mathbf{L}}\left(i\bm{k}\right)\Delta t/2}\mathcal{P}\left(\bm{k}\right)\mathcal{F}$. 

A common choice is a projector $\mathcal{P}\left(\bm{k}\right)$ such
that, for a maximum momentum of $k_{m}=\pi/\Delta x$ and a lattice
step-size of $\Delta x$:
\begin{equation}
\mathcal{P}\left(\bm{k}\right)=\begin{cases}
0 & \mathrm{for}\ \left|k_{i}\right|>k_{m}/2,\\
1 & \mathrm{for}\ \left|k_{i}\right|<k_{m}/2.
\end{cases}
\end{equation}
This removes aliasing errors \citep{orszag1971,Derevyanko2008Rule},
which are well-known in many applications of Fourier transforms \citep{boyd2001chebyshev}.
More sophisticated algorithms are also possible, that either remove
high momentum components more gradually \citep{DRUMMOND1983211},
or else require less storage by using a phase-shift technique developed
for equations in hydrodynamic studies of turbulence, that cancels
only the aliased modes \citep{rogallo1981numerical}. 

\subsection{Complex absorption model}

The apodisation behavior required to prevent aliasing in space is
different to momentum, because of the different effects of the equations,
which produce a localized spread in position from second-order derivative
terms. Therefore we consider a generic apodisation as a space dependent
absorbing reservoir, which includes a complex absorption coefficient.
This modifies the original field equations, so that

\begin{align}
\dot{\psi}_{j}(\bm{x}) & =\mathcal{D}_{j}\left[\mathbf{\overrightarrow{\psi}}\right]-\gamma_{j}(\bm{x})\psi_{j}(\bm{x}).
\end{align}

We will start by considering classical apodisation, and include the
effects of quantum noise in subsection (\ref{subsec:Quantum-noise-terms}).
If applied to a periodic calculation with a transverse lattice spacing
of $\Delta x$ and boundaries at $x_{m}=L/2$, we require that there
is no absorption at the center, so that $\gamma(0)=0\,,$ and $\gamma(\pm x_{m})\rightarrow\infty$
at the boundaries. However, it is sufficient to simply have a very
high absorption at the boundary, rather than an infinite absorption. 

The graphs given below are of the number density $\rho_{i}=\left|\psi_{i}\right|^{2}$,
where Fig (1) and (2) show the errors occurring in $\rho_{1}=\left|\psi_{1}\right|^{2}$
in a non-apodised case. The first figure has a spatial domain of $x=\pm10$.
The central error for Gaussian diffraction in a doubled space domain
of of $x=\pm20$ is shown in Fig (2). Errors at $\left|x\right|=10$
are of order of $3\%$ of the initial on-axis intensity despite the
enlarged boundaries. This shows that the aliasing errors can only
be removed slowly by increasing the spatial volume. 

All numerical examples obtained here used a central difference Fourier
transform algorithm \citep{Werner:1997}, with public-domain stochastic
partial differential equation software \citep{kiesewetter2016xspde}.

\begin{figure}
\includegraphics[width=0.75\columnwidth]{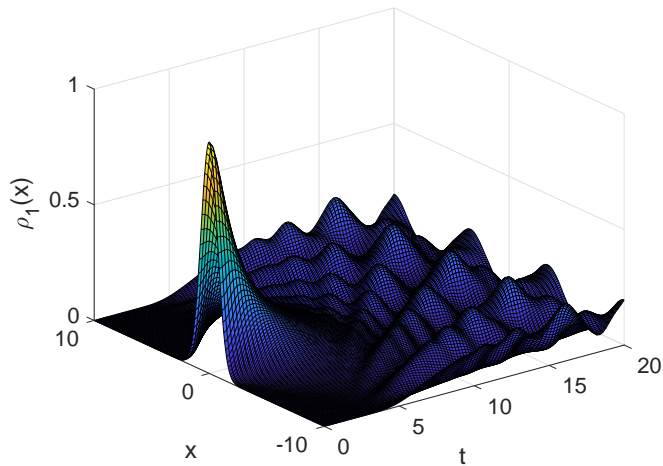}\caption{Non-apodised diffraction, Eq (\ref{eq:diffraction}), in a finite
interval with periodic boundary conditions. Graph is of $\left|\psi\right|^{2}$
versus $x$, $t$ for a Gaussian of initial width $\sigma=1$, integrated
out to $t=20$ in a spatial region of $\pm10$. Step sizes were $\Delta t=0.025$
and $\Delta x=0.078$.}
\end{figure}

To achieve a minimum error in the central propagation region, the
form considered here for apodisation in space is complex, so that
it can be expressed in terms of real fields, $\gamma'(\bm{x})$ and
$V(\bm{x})$, so that
\begin{equation}
\gamma(\bm{x})=\gamma'(\bm{x})+iV(\bm{x}).
\end{equation}
This complex apodisation, including phase-shifts as well as absorption,
is then expanded as a polynomial in the transverse coordinates, where
we consider just a single transverse dimension for simplicity:

\begin{equation}
\gamma(\bm{x})=\sum_{q=0}^{p}\gamma_{q}x^{q}.
\end{equation}

This allows one to investigate a range of possible models of apodisation.
Allowing for both real and complex terms corresponds to including
arbitrary polynomial potentials $V(\bm{x})$ to shape the solution,
in addition to the absorption term, $\gamma'$. These correspond physically
to a change in refractive index in optics, or an external potential
in quantum mechanics. The challenge is to find an absorption function
that minimizes errors caused by the periodic boundaries, and gives
a result for r $\left|x\right|<x_{m}/2$ close to the exact one. 

\begin{figure}
\includegraphics[width=0.75\columnwidth]{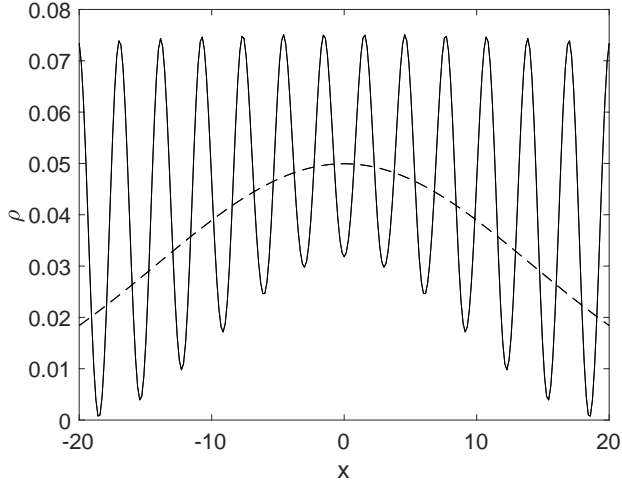}\caption{Non-apodised diffraction, Eq (\ref{eq:diffraction}), in a larger
interval with periodic boundary conditions (solid line), compared
to the exact result (dashed line). Graph is of $\left|\psi\right|^{2}$
versus $x$, at $t=20$ for a Gaussian of initial width $\sigma=1$,
in a spatial region of $\pm20$. Step sizes were $\Delta t=0.005$
and $\Delta x=0.157$. }
\end{figure}

\subsection{Time-evolution}

To illustrate this approach, we analyze a relatively simple case in
one dimension of the paraxial equation for diffraction or dispersion
of a field. To treat this problem, we use an analytic theory. Suppose
that there is an evolution equation in an infinite spatial region,
which is scaled to have the form of:
\begin{equation}
\partial_{t}\psi=\frac{i}{2}\partial_{x}^{2}\psi.\label{eq:diffraction}
\end{equation}
The apodisation method involves adding an absorption $\gamma\left(t,x\right)$,
so that:
\begin{equation}
\partial_{t}\psi=\frac{i}{2}\partial_{x}^{2}\psi-\gamma\psi.
\end{equation}
Here the absorption term $\gamma\left(t,x\right)$ is chosen so that
the solution to the modified equation can be treated in a finite region
with periodic boundary conditions, having a minimum error relative
to the exact solution in the central region, in this case for $\left|x\right|<x_{m}/2$.

To analyze the problem, consider a solution in one transverse dimension
using a power series expansion for $\log\psi$, where only even terms
are assumed. There is an overall negative sign to ensure that coefficients
$\alpha_{p}$ are positive for bounded solutions: 
\begin{equation}
\psi=\exp\left(-\sum_{q}\alpha_{q}\left(t\right)x^{2q}\right).\label{eq:psi-series}
\end{equation}
The expansion coefficients $\alpha_{p}\left(t\right)$ must satisfy
a recursion relation.  As a result, the expansion coefficients evolve
in time according to: 

\[
\dot{\alpha}_{q}=\gamma_{q}-i\beta_{q}+\frac{i}{2}\left(2q+2\right)\left(2q+1\right)\alpha_{q+1}\,,
\]
where the nonlinear term coming from the second derivative is expanded
as:

\begin{align}
\frac{1}{2}\left(\sum_{q}2q\alpha_{q}x^{2q-1}\right)^{2} & =\sum_{q}\beta_{q}x^{2q}.
\end{align}

Here, $\beta_{1}=2\alpha_{1}^{2}$, $\beta_{2}=8\alpha_{1}\alpha_{2}$
, $\beta_{3}=12\alpha_{1}\alpha_{3}+8\alpha_{2}^{2}$, $\beta_{4}=\left(16\alpha_{1}\alpha_{4}+24\alpha_{2}\alpha_{3}\right),$
and so on. The nonlinear coefficients $\beta_{q}$ have the universal
property that the highest order term appearing for $q>1$ is $4q\alpha_{q}\alpha_{1}$,
and all other terms have lower orders in $\alpha_{q}$. 

The solution to the apodisation problem depends on the exact solution
that it is intended to approximate. This actually makes it highly
nontrivial. Here we apodise an exactly known Gaussian diffraction
problem. Such cases have behavior near the boundaries that resembles
the unknown solution of interest. Even though the real problem typically
has additional nonlinear terms in its time-evolution equation, the
extra nonlinear terms are negligible at low intensities near boundaries.

Suppose that only $\alpha_{0}$ and $\alpha_{1}$ are nonzero without
apodisation, so that the exact Gaussian solution has: 
\begin{align}
\dot{\alpha}_{0} & =i\alpha_{1}\nonumber \\
\dot{\alpha}_{1} & =-2i\alpha_{1}^{2}.
\end{align}
This provides a reference solution for the non-apodised case. Initially,
let $\alpha_{1}\left(0\right)=1/\left(2\sigma^{2}\right)$, where
$\sigma$ is a scale length that defines the size of the Gaussian
beam. It is possible to choose an apodisation strategy that is independent
of $\sigma$, provided there is a large spatial window. With this
choice, one obtains:

\begin{equation}
\alpha_{1}\left(t\right)=\frac{1}{2\left(\sigma^{2}+it\right)},
\end{equation}
and therefore one obtains the zero-th order solution, as expected,
of
\begin{align}
\alpha_{0} & =C+\frac{1}{2}\ln\left(\sigma^{2}+it\right).
\end{align}

This is a standard Gaussian diffraction/dispersion solution, which
has the property that it spreads out indefinitely. The result is that
the central intensity tends to zero at large $t$, since $\psi\left(t,0\right)\sim1/\sqrt{t-i\sigma^{2}}$.
If the same equations are solved with periodic boundary conditions,
the solution has a conserved total number, and simply becomes a mass
of interference fringes as shown in Fig (1). These fringes only decay
slowly with increasing boundary size, as shown in Fig (2). 

To treat apodisation, one may consider terms up to $p-th$ order,
for some maximum order $p$, and we define $\Gamma_{p}=\gamma_{p}x_{m}^{p}$
as the boundary absorption. We suppose that $\alpha_{0},\alpha_{1}$
are nonzero initially. To give a definite example, suppose that $\alpha_{q}=0$
initially for $q>1$, as in the exact solution given above. The low-order
terms are as usual for a Gaussian, and there also must be an envelope
$\alpha_{p}$ in the apodised solution, which prevents wrap-around.
This is obtained through adding apodizing terms such that $\alpha_{q}=0$
for $p>q>1$. Fig (3) shows the result of purely absorbing apodisation,
with one term of $\Gamma_{10}=10$, giving reduced errors of order
of $7\times10^{-5}$ of the initial on-axis intensity.

\begin{figure}
\includegraphics[width=0.75\columnwidth]{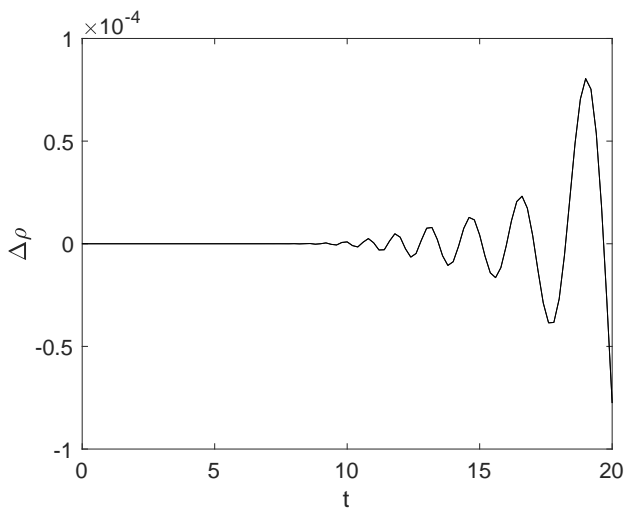}\caption{Apodised diffraction in a finite interval with periodic boundary conditions,
using tenth order apodisation, with $\Gamma_{10}=10$, and no additional
complex phase-shift. Graph is of central error in intensity versus
$t$. All integration parameters as in Fig (2). }
\end{figure}

\subsection{Complex apodisation}

We treat a $p$-th order apodisation. From matching powers in the
two series, there are $1+p$ corresponding time-evolution equations.The
higher-order terms for $q>p$ are assumed to have relatively small
magnitudes, and are neglected. We wish to specify the apodisation
so that $\alpha_{2}$ and $\alpha_{0}$ are unchanged from their free-field
behavior given above, to obtain behavior near the center that is invariant
under apodisation, with all the other terms zero up to $\alpha_{p}$.
This implies that $\gamma_{q}=0$ for $0\le q<p-1$, and
\begin{equation}
0=\gamma_{p-1}+ip\left(2p-1\right)\alpha_{p}.
\end{equation}

It is essential that $\alpha_{p}$ has a positive real part to apodise
the wave-function as it spreads near the boundaries. Hence, $\gamma_{p-1}$
should have complex values. This additional quartic potential with
$V_{q-1}<0$ provides a linear phase-shift that cancels diffractive
shifts induced by the apodisation.

As a result, since the time-evolution of $\alpha_{1}$ is known, the
$p$-th term must satisfy:
\begin{align}
\dot{\alpha}_{p} & =\frac{-2p\alpha_{p}}{t-i\sigma^{2}}+\gamma_{p}.
\end{align}
To obtain an analytic solution, we first consider a constant apodisation
$\gamma_{p}$, and introduce a new time variable $\tau=\ln\left(t-i\sigma^{2}\right)$.
The amplitude equation then becomes: 

\begin{equation}
\frac{\partial\alpha_{p}}{\partial\tau}=-2p\alpha_{p}+\gamma_{p}e^{\tau}.
\end{equation}

On solving this equation, one finds that the p-th order amplitude
has a growing behavior in time,which has a well-defined limit after
transforming back to the original time variables, such that: 
\begin{equation}
\lim_{t\rightarrow\infty}\alpha_{p}=\frac{\gamma_{p}}{2p+1}\left(t-i\sigma^{2}\right).
\end{equation}
Therefore, to prevent possible lower-order amplitude changes, one
should include a time-dependent term or order $p-1,$ such that:
\begin{equation}
0=\gamma_{p-1}+ip\left(2p-1\right)\alpha_{p}.
\end{equation}
 At long times, one finds that
\begin{equation}
\gamma_{p-1}=\frac{-ip\left(2p-1\right)\gamma_{p}}{2p+1}\left(t-i\sigma^{2}\right).
\end{equation}

There is some growth in the truncation error with time, since $\alpha_{p}$
increases linearly. This is not as severe a problem as one might imagine.
The growth simply means that there is a truncation near the boundaries,
just as one might expect from having an absorbing region. This also
leads to an apparent non-conservation of energy. Again, this is physically
reasonable, as diffraction leads to escape through the boundaries,
which can be monitored numerically as explained below.

In order to understand the scaling with window size, suppose that
$\gamma_{p}$ is defined by the maximum absorption $\Gamma_{p}$ at
the window boundary of $x=\pm x_{m}$, so that
\begin{equation}
\gamma_{p}=x_{m}^{-2p}\Gamma_{p}.
\end{equation}

The maximum size of the optimal complex term, $\gamma_{p-1}$, reduces
as the window size increases, with an asymptotic behavior in time
that is purely imaginary and independent of the initial Gaussian width:
\begin{equation}
\gamma_{p-1}=\frac{-ip\left(2p-1\right)}{2p+1}\frac{t}{x_{m}^{2p}}\Gamma_{p}.\label{eq:complex phase-shift}
\end{equation}

\begin{figure}
\includegraphics[width=0.75\columnwidth]{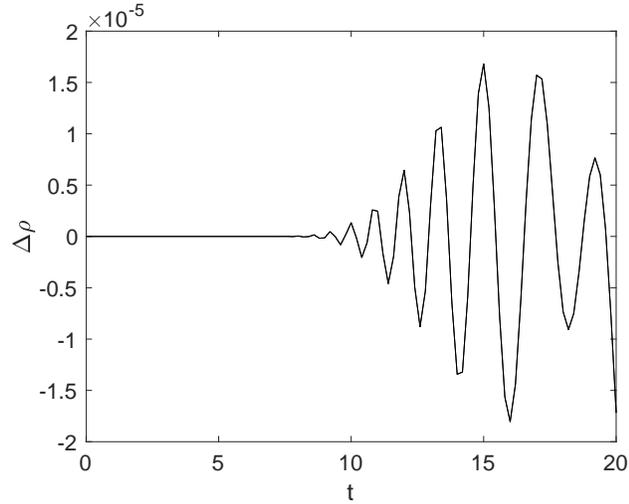}\caption{Apodised diffraction in a finite interval with periodic boundary conditions,
using tenth order apodisation, with $\Gamma_{10}=10$, and a complex
phase-shift as in Eq (\ref{eq:complex phase-shift}). Graph is of
central error in intensity versus $t$. All integration parameters
as in Fig (2). }
\end{figure}
As a result, a feasible strategy of checking the boundary error is
to increase the window size with constant $\Gamma_{p}$, where $\Gamma_{p}$
is the boundary value of the highest order apodisation term. This
should result in negligible change in the central region. The required
asymptotic form is independent of the scale factor $\sigma$, so that
this phase-shift technique can be used even when the correct scale
factor is only known approximately, as it is for non-Gaussian and
nonlinear calculations.

\begin{figure}
\includegraphics[width=0.75\columnwidth]{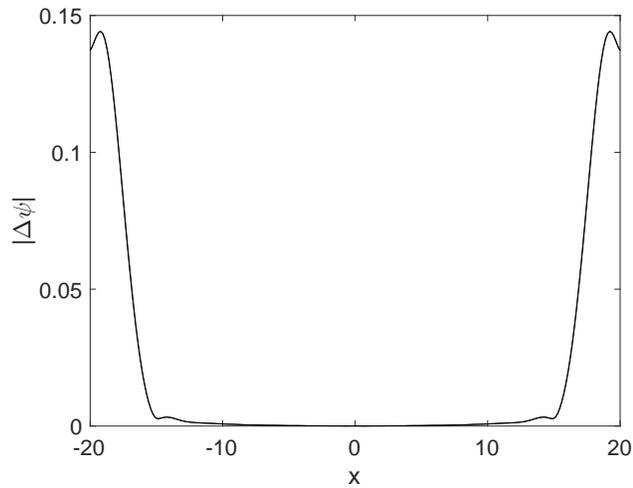}\caption{Apodised diffraction in a finite interval with periodic boundary conditions,
using tenth order apodisation, with $\Gamma_{10}=10$, and a complex
phase-shift as in Eq (\ref{eq:complex phase-shift}). Graph is of
wave-function errors $\left|\Delta\psi\right|=\left|\psi_{a}-\psi_{exact}\right|$
versus $x$ at $t=20$. Here the apodised wavefunction is $\psi_{a}$.
All integration parameters as in Fig (2). }
\end{figure}

Results are shown in Fig(4), with intensity errors reduced to less
than $2\times10^{-5}$ of the initial on-axis intensity, and in Fig
(5), which shows that wave-function errors are reduced to less than
$7\times10^{-4}$ across the entire region of $\left|x\right|<10$. 

A constant apodisation leads to a growing error in $\alpha_{p}$.
It is also possible to consider a time-dependent apodisation. This
leads to solutions where $\alpha_{p}$ is asymptotically constant.
However, this is not investigated here. Such methods appear to be
closely dependent on the exact solution that is targeted. Hence, they
may be less useful generally.

\subsection{\label{subsec:Quantum-noise-terms}Quantum noise terms and number
conservation}

Because of commutation relations, a quantum apodisation must be accompanied
by noise terms, in either the Wigner or Q methods. This is not needed
for P-distributions due to normal-ordering. Such terms are also not
needed with projective anti-aliasing in momentum space, since the
amplitude of the projected high-momentum modes is maintained at zero.
In $s-$ordered simulations, apodizing absorption should be accompanied
by noise, so that:
\begin{equation}
\dot{\psi}=-\gamma(\bm{x})\psi+\sqrt{\gamma'(\bm{x})}\zeta\left(\bm{x}\right)\label{eq:apodise_noise}
\end{equation}
where we omit
\begin{equation}
\left\langle \zeta\left(\bm{x}\right)\zeta^{*}\left(\bm{x}'\right)\right\rangle =2s\delta\left(\bm{x}-\bm{x}'\right)
\end{equation}
 The purpose of the noise is to ensure that the $s-$ordered quantum
fields have a vacuum occupation of $s$ bosons per mode, with

\begin{equation}
\left\langle \psi\left(\bm{x}\right)\psi^{*}\left(\bm{x}'\right)\right\rangle _{vac}=s\delta\left(\bm{x}-\bm{x}'\right).
\end{equation}

Effectively, the apodisation procedure adds an absorbing zero-temperature
reservoir that prevents particles from reaching the boundary. Suppose
one wishes to record how many particles have been absorbed? In this
case, a second, reservoir occupation field is needed, as shown in
Fig (6) for the test case of Gaussian diffraction. 
\begin{figure}
\includegraphics[width=0.75\columnwidth]{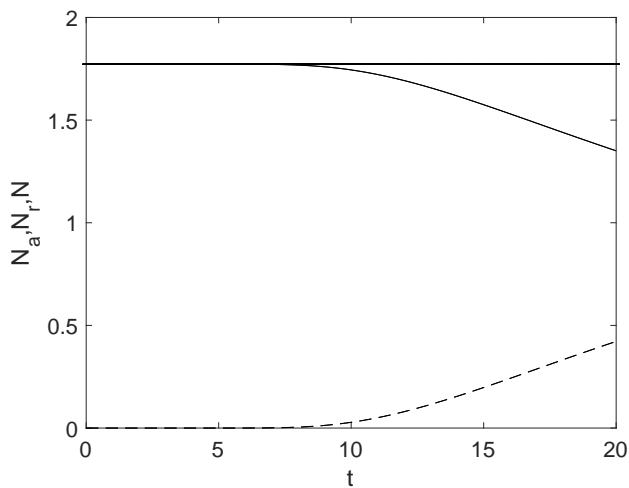}\caption{Apodised diffraction in a finite interval with periodic boundary conditions.
Graph is of apodised number $N_{a}$ (solid line), reservoir number
$N_{r}$ (dashed line), and the conserved total number (horizontal
black line). Other parameters as in Fig(4).}
\end{figure}

The $s-$ordered quantum apodisation equations (\ref{eq:apodise_noise})
have the property that, if we define 
\begin{equation}
\rho_{\psi}\left(\bm{x}\right)=\psi\left(\bm{x}\right)\psi^{*}\left(\bm{x}\right)
\end{equation}
then the expectation values change in time due to the reservoirs so
that:
\begin{equation}
\left\langle \dot{\rho}_{1}\left(\bm{x}\right)\right\rangle =2\gamma'(\bm{x})\left[\frac{s}{\Delta V}-\left\langle \rho_{\psi}\left(\bm{x}\right)\right\rangle \right]
\end{equation}

Hence, we can add a reservoir number field that reflects the complementary
process in which atoms are transferred to the reservoir while vacuum
fluctuations are lost, which satisfies the following Stratonovich
stochastic equation:
\begin{equation}
\dot{\rho}_{2}=2\gamma'(\bm{x})\psi\psi^{*}-\left(\sqrt{\gamma'(\bm{x})}\zeta\left(\bm{x}\right)\psi^{*}+h.c\right)
\end{equation}

As a result, provided there is no other reservoir present, the total
particle number is conserved:
\begin{equation}
\frac{d}{dt}\left\langle \rho_{1}\left(\bm{x}\right)+\rho_{2}\left(\bm{x}\right)\right\rangle =0.
\end{equation}
This allows the total number of particles to be tracked, independent
of boundary absorption and apodisation.

\section{Conclusions}

This paper discusses ``quantum software'' applied to the simulation
of quantum field propagation with $s-$ordered phase-space methods.
Typically, via probabilistic sampling, one obtains stochastic partial
differential equations which can be solved numerically even in Hilbert
spaces of large size. There are other, more subtle issues that arise
during time-evolution. These are the generation of arbitrary initial
quantum states, and the treatment of finite spatial boundaries.

For initial Gaussian quantum states, the initial sampling is straightforward.
Complex weighting techniques are more efficient when the initial states
are number states. This provides a way for sampling distributions
for any quantum states with arbitrary ordering. Similarly, we show
that complex apodisation methods can provide an efficient means of
treating infinite domains using Fourier transform methods. However,
in such cases, noise terms are needed to ensure that commutation relations
are respected. 

\section*{Acknowledgements}

Discussions with Bogdan Opanchuk, Boris Malomed, Vladimir Yurovsky,
King Ng, Run Yan Teh and Maxim Olshanii are gratefully acknowledged.
This work was funded by an Australian Research Council Discovery Grant,
and by the Joint Institute for Laboratory Astrophysics through their
Visiting Fellowship program.

\bibliographystyle{apsrev4-1}
\bibliography{RMP_Draft_references}

\end{document}